\documentclass[aps,prl,twocolumn,amsmath,amssymb,nofootinbib,superscriptaddress,floatfix]{revtex4}

\usepackage{bm} 
\usepackage{graphicx}
\usepackage{amssymb}
\usepackage{curves}
\usepackage{amsmath}
\usepackage{fancybox}
\usepackage{dcolumn} 
\usepackage[dvips]{color}
\usepackage{epic}
\usepackage{longtable}

\begin{document}

\title{
$\mathbb{Z}^{\ }_2$ topological term, the global anomaly,
and the two-dimensional symplectic symmetry class 
of Anderson localization
 }

\author{Shinsei Ryu}
\affiliation{Kavli Institute for Theoretical Physics,
 	     University of California, 
 	     Santa Barbara, 
 	     CA 93106, 
 	     USA}
\author{Christopher Mudry}
\affiliation{Condensed Matter Theory Group,
              Paul Scherrer Institute,
              CH-5232 Villigen PSI,
 	      Switzerland}

\author{Hideaki Obuse}
\affiliation{Condensed Matter Theory Laboratory,
             RIKEN, 
             Wako, 
             Saitama 351-0198, 
             Japan}

\author{Akira Furusaki}
\affiliation{Condensed Matter Theory Laboratory,
             RIKEN, 
             Wako, 
             Saitama 351-0198, 
             Japan}

\date{\today}

\begin{abstract}
We discuss,
for a two-dimensional Dirac Hamiltonian 
with random scalar potential,
the presence of
a $\mathbb{Z}^{\ }_2$ topological term 
in the non-linear sigma model encoding
the physics of Anderson localization in the symplectic symmetry class.
The $\mathbb{Z}^{\ }_2$ topological term realizes 
the sign of the Pfaffian of a family of Dirac operators.
We compute the corresponding global anomaly, i.e.,
the change in the sign of the Pfaffian 
by studying a spectral flow numerically.
This $\mathbb{Z}^{\ }_2$ topological effect
can be relevant to graphene when the impurity potential is long-ranged
and, also, to the two-dimensional boundaries of a three-dimensional 
lattice model of $\mathbb{Z}^{\ }_2$ 
topological insulators in the symplectic symmetry class.
\end{abstract}

\maketitle

There is strong theoretical and numerical supporting evidence
for the premise that metal-insulator transitions in the problem of
Anderson localization with short-range correlated disorder
can be classified in terms of a 
few fundamental properties of microscopic Hamiltonians. 
Whenever the motion of the relevant (fermionic) quasiparticles
is diffusive, a non-linear sigma model (NL$\sigma$M)
can be derived for each symmetry class.
A NL$\sigma$M encodes 
fluctuations of Nambu-Goldstone bosons 
(diffuson and Cooperons)
that are defined on a curved manifold 
(target manifold) determined by
the symmetries of a microscopic disordered Hamiltonian.

When the fluctuations are small,
the dynamics of the Nambu-Goldstone bosons
are solely determined by the local data of the target manifold
such as the metric and curvature.
However, the global topology of the target manifold
does have important effects.
For example,
a weak magnetic field relative to the disorder strength
leads to localization of all states 
in two-dimensional space.
On the other hand, a strong magnetic field leads to 
the integer quantum Hall plateau transition.
This difference is captured by the absence or presence,
respectively, of a topological term in the NL$\sigma$M 
corresponding to the two-dimensional unitary symmetry class \cite{Pruisken84}.
Similarly, a random single-particle Hamiltonian 
that preserves time-reversal symmetry but breaks spin-rotation
symmetry defines the symplectic symmetry class and can be associated
to a NL$\sigma$M. In two-dimensional space, Fendley
\cite{Fendley01}
made the observation that such a NL$\sigma$M admits a topological term.
The configuration space of 
the relevant NL$\sigma$M has two topologically distinct sectors. 
Configurations from different topological sectors can thus 
be given different Boltzmann weights in the presence of the topological term, 
thereby realizing the outcome of a 
metal-insulator transition belonging to a
universality class different from the one without
the topological term \cite{Fendley01}. 

We show in this letter that a topological term,
encoding a global (i.e., nonperturbative) anomaly, 
is realized by the Pfaffian of Majorana fermions 
that originate from the problem of
Anderson localization defined by the two-dimensional
two-component Dirac Hamiltonian
\begin{equation}
\mathcal{H}=
\bm{\sigma}\cdot\bm{p}
+
\sigma^{\ }_0\,V(\bm{r}),
\qquad
\bm{p}:=
\frac{\partial}{{i}\partial\bm{r}},
\label{eq: single dirac hamiltonian}
\end{equation}
subject to a white-noise and Gaussian
correlated random scalar potential
$V(\bm{r})$,
$
\overline{V(\bm{r})}=0$,
$\overline{V(\bm{r})V(\bm{r}')}=
g\,
\delta(\bm{r}-\bm{r}')$.
Here, 
$\bm{r}\in\mathbb{R}^2$, $g\geq0$ measures the disorder strength, 
and $\sigma^{\ }_{x,y,z}$ denote the standard Pauli matrices
with $\sigma^{\ }_{0} $ the corresponding unit matrix.
In the recent independent derivation 
by Ostrovsky et al.\ \cite{Ostrovsky07},
the topological term is not presented in the form of the sign
of a Pfaffian which, however, is essential for it to be 
interpreted as a global anomaly.
The fabrication of graphene 
\cite{Novoselov04}
has triggered a renewed theoretical interest in the properties
of the random Dirac Hamiltonian%
~(\ref{eq: single dirac hamiltonian}).
We will argue that the random Dirac Hamiltonian%
~(\ref{eq: single dirac hamiltonian})
can be realized at the two-dimensional boundaries 
and in the low-energy limit of three-dimensional lattice models.

As observed by Ludwig et al.\ \cite{Ludwig94} the symmetry
$
{i}\sigma_y
\mathcal{H}^*
\left( -{i}\sigma_y \right)
=
\mathcal{H},
$
which we shall abusively call time-reversal symmetry (TRS) 
as it might not necessarily realize the TRS
of the underlying microscopic model,
puts the random Dirac Hamiltonian~(\ref{eq: single dirac hamiltonian})
in the two-dimensional symplectic symmetry class.
The disorder in the Dirac Hamiltonian~(\ref{eq: single dirac hamiltonian})
is thus expected to yield weak anti-localization corrections to the mean conductance
\cite{Ando98}.

The link between single-particle random Hamiltonians
and NL$\sigma$M comes about when setting up a generating
function for the mean value taken by the product of the
Green's functions.
Using the standard machinery of the replica trick \cite{Efetov80}, 
the disorder-averaged product of the retarded and
advanced Green's functions of Eq.~(\ref{eq: single dirac hamiltonian})
can be obtained from the Grassmann path integral
$\overline{Z}:=
\int \mathcal{D}\left[\bar\psi,\psi \right] 
\exp\left(
-\int d^2 r\, \mathcal{L}\right),
$
where
\begin{equation}
\mathcal{L}=
\bar{\psi}^{\ }_a
\left(
\bm{\sigma}\cdot\bm{p}\,
\delta^{\ }_{ab}
+
{i}
\eta \Lambda^{\ }_{ab}
\right)
\psi^{\ }_b
-
\frac{g}{2}
\bar{\psi}^{\ }_a\psi^{\ }_a
\bar{\psi}^{\ }_b\psi^{\ }_b,
\label{eq: interacting fermionic effective action}
\end{equation}
and with the Majorana condition
$ 
\bar{\psi}^{\ }_{a}:=
\psi^{T}_{a}
{i}\sigma^{\ }_y,
$
on the Grassmann integration variables
($a=1,\ldots,4N^{\ }_{\mathrm{r}}$).
Summation over repeated 
indices with $a,b=1,\ldots,4N^{\ }_{\mathrm{r}}$
is implied. For any $a=1,\ldots,4N^{\ }_{\mathrm{r}}$,
the Pauli matrices 
act on the two components of the spinor $\psi^{\ }_{a}$.
The 
$4N^{\ }_{\mathrm{r}}\times4N^{\ }_{\mathrm{r}}$
diagonal matrix 
$\Lambda = \mathrm{diag}\,
\left( \mathbb{I}_{2N_{\mathrm{r}}},
-\mathbb{I}_{2N_{\mathrm{r}}}
\right)
$
distinguishes 
two sectors coming from
the retarded and advanced Green's functions. 
It is multiplied by the infinitesimal
positive number $\eta$. 
The replica limit $N^{\ }_{\mathrm{r}}\to0$ 
is understood if the effective partition function is
purely fermionic, while the choice $N^{\ }_{\mathrm{r}}=1$
is appropriate when 
(\ref{eq: interacting fermionic effective action})
is regarded as the fermionic sector of 
a supersymmetric representation 
of the product of the retarded and advanced
Green's functions.
The 4-fermion interaction
in (\ref{eq: interacting fermionic effective action})
originates from the integration over the static random field
$V(\bm{r})$.

When $\eta=0$,
the replicated action is invariant under 
a $\mathrm{O}(4N^{\ }_\mathrm{r})$ transformation
acting on the
$(\mbox{time-reversal}) \otimes
( \mbox{retarded/advanced}) \otimes (\mbox{replica})$
indices.
The small imaginary part of the
energy $\eta$ lifts the degeneracy
between the retarded and advanced sectors, thereby
reducing the $\mathrm{O}(4N^{\ }_{\mathrm{r}})$
symmetry down to $\mathrm{O}(2N^{\ }_{\mathrm{r}})
\times \mathrm{O}(2N^{\ }_{\mathrm{r}})$.
The Nambu-Goldstone modes
associated with this symmetry breaking are
described by a bosonic matrix field $Q(\bm{r})$ living
in the coset space 
$G/H\equiv 
\mathrm{O}(4N^{\ }_{\mathrm{r}})/
\mathrm{O}(2N^{\ }_{\mathrm{r}})\times
\mathrm{O}(2N^{\ }_{\mathrm{r}})
$. 
They emerge after performing
the Hubbard-Stratonovich decoupling
of the interaction 
by a $\mathrm{O}(4N^{\ }_{\mathrm{r}})$ matrix field
and then freezing its massive modes
(i.e., all modes other than the Nambu-Goldstone modes).
Their interactions are governed by the effective local field theory 
\begin{subequations}
\begin{equation}
\begin{split}
&
\overline{Z}_{\mathrm{eff}}=
\int 
\mathcal{D}\left[\bar{\psi},\psi\right] 
\mathcal{D}[Q]\
e^{
-\int d^2 r
\left[
\mathcal{L}^{\ }_{\mathrm{f}}
+
\frac{\Delta^2}{4g}
\mathrm{tr}\,
\left(Q Q^T \right)
\right]
,
}
\\
&
\mathcal{L}^{\ }_{\mathrm{f}}
:=
\bar{\psi}\, 
D[Q]\, 
\psi
=
\psi^T\,
\widetilde{D}[Q] \,
\psi.
\end{split}
\label{eq: def eff action}
\end{equation}
Here, we have introduced the Majorana kernel
\begin{equation}
\begin{split}
\left(\widetilde{D}[Q]\right)^{\ }_{ab}:=&\,
{i}\sigma^{\ }_{y}
\left(
\bm{\sigma}\cdot\bm{p}\,
\delta^{\ }_{ab}
-
{i}
\Delta Q^{\ }_{ab}
+
{i}\eta
\Lambda^{\ }_{ab}
\right)
\\
=:&\,
{i}\sigma^{\ }_{y}
\left(D[Q]\right)^{\ }_{ab},
\quad
a,b=1,\ldots,4N^{\ }_{\mathrm{r}}.
\end{split}
\label{eq: def Majorana widetilde D}
\end{equation} 
It is skew symmetric,
$
\left(
{i}\sigma^{\ }_y D[Q]
\right)^{T}=
-
{i}\sigma^{\ }_y D[Q],
$
in the background of $Q\in G/H$, i.e., 
\begin{eqnarray}
Q^{2}=\mathbb{I}^{\ }_{4N^{\ }_{\mathrm{r}}}, 
\quad 
Q=Q^T, 
\quad 
\mathrm{tr}\,Q =0.
\end{eqnarray}
\end{subequations}
The real number $\Delta$,
which represents the imaginary part of the self-energy
caused by disorder,  
is determined by solving the self-consistent equation
$
\pi/g
=
\ln
\left[ 
1
+
(
a^{\ }_0 \Delta
)^{-2} 
\right]
$
where $a^{\ }_0$ is the short-distance cutoff.

For smooth spatial variations of the $Q$-field 
starting from $\Lambda$,
the spectrum of $\widetilde{D}[Q]$
has a gap of order $\Delta$ at the band center.
We can then ignore $\eta$ and integrate the Majorana fermions out. 
An effective action for the Nambu-Goldstone modes follows.
This integration, when properly regularized, 
\textit{defines} the Pfaffian 
$
\int 
\mathcal{D}\left[\bar{\psi},\psi\right]
\exp\left(
-\int d^2 r\, \mathcal{L}^{\ }_{\mathrm{f}} \right)
\equiv
\mathrm{Pf}
\left(
\widetilde{D}[Q]
\right).
$
A possible regularization that preserves TRS 
starts with wrapping momentum space around the two-torus $T^{2}$.
Momentum is then quantized,
$k^{\ }_{\mu}= 2\pi n^{\ }_{\mu}/L$ where $\mu=1,2$,
with the level spacing $2\pi/L$ controlled by the long-distance cutoff $L$.
Momentum is made countably finite,
$n^{\ }_{\mu}= -(N-1)/2,-(N-1)/2+1,\ldots,(N-1)/2$ where $\mu=1,2$
with the help of the short-distance cutoff $a^{\ }_{0}$.
The ratio $N=L/a^{\ }_{0}$ of the ultraviolet to infrared cutoff
is chosen an odd integer for convenience.
To complete the definition of the regularized 
Pfaffian we note that its square is the determinant
$
\mathrm{Det}\,
\widetilde{D}[Q]
$.
This determinant can be rewritten
\begin{subequations}
\label{eq: key identity between det}
\begin{equation}
\mathrm{Det}\,
\widetilde{D}[Q]
=
\mathrm{Det}\,
\left(
{i}\sigma^{\ }_{y}
D'[Q]
\right)
=
\mathrm{Det}\,
D'[Q]
\label{eq: Det D'}
\end{equation}
with the Dirac kernel
($a,b=1,\ldots,4N^{\ }_{\mathrm{r}}$)
\begin{equation}
\left(
D'[Q]
\right)^{\ }_{ab}
=
\bm{\sigma}\cdot\bm{p}\,
\delta^{\ }_{ab}
+
\sigma^{\ }_z\,\Delta\,Q^{\ }_{ab}.
\label{eq: Dirac Kernel D'}
\end{equation}
\end{subequations}
The real-valued spectrum of $D'[Q]$ comes in pairs
of non-vanishing eigenvalues
$\{-\lambda^{\prime}_i,+\lambda^{\prime}_i\}$
labeled by the index $i$ running over some
countably finite set.
For some reference configuration 
$Q$,
we \textit{define}
\begin{equation}
\mathrm{Pf}\,
\widetilde{D}[Q] 
:=
\prod_i \widetilde{\lambda}^{\ }_i,
\label{eq: def Pf of Q}
\end{equation}
where $\widetilde{\lambda}^{\ }_i\in\mathbb{R}$ represents 
either one of $-\lambda^{\prime}_i$ or $+\lambda^{\prime}_i$ for each $i$.
It follows that
$
\mathrm{arg}\,
\mathrm{Pf}\,
\widetilde{D}[Q] 
\in
\{0,\pi\}.
$
The sign of the Pfaffian is protected by the spectral gap
$\propto\Delta$ under any infinitesimal change of $Q$.
This protection does not extend to all configurations.
Indeed, the second homotopy group 
of the target manifold is non-trivial,
\begin{equation}
\pi^{\ }_2(G/H)=
\left\{
\begin{array}{ll}
\mathbb{Z}^{\ }_2,
&
\hbox{ for all $N^{\ }_{\mathrm{r}}>1$,}
\\
&
\\
\mathbb{Z}\times \mathbb{Z},
&
\hbox{ for $N^{\ }_{\mathrm{r}}=1$,}
\end{array}
\right.
\label{eq: second homotopy group}
\end{equation}
and, as we are going to show explicitly,
there is a phase difference of $\pi$ between
the two Pfaffians~(\ref{eq: def Pf of Q})
when evaluated at two
$Q$'s belonging to distinct $\mathbb{Z}^{\ }_{2}$
topological sectors.
This ambiguity in fixing the sign of the
Pfaffian over all configurations,
starting from some reference one,
is reminiscent of the SU(2) global anomaly in four space-time dimensions
that follows from 
$\pi^{\ }_4(\mathrm{SU}(2))=\mathbb{Z}^{\ }_2$
\cite{Witten82}.

We turn to the construction of two families of $Q$-fields that differ
in the sign of the Pfaffian~(\ref{eq: def Pf of Q}). 
To this end, we first compactify
two-dimensional Euclidean space $\bm{r}\in\mathbb{R}^{2}$
by wrapping it once around the two-sphere $S^{2}$. On the two-sphere
parametrized by the 
polar $-\pi/2\leq\theta\leq+\pi/2$ 
and azimuthal $0\leq\phi<2\pi$ angles,
we define, following Weinberg et al.\ \cite{Weinberg84}, 
the family of fields
\begin{equation}
\begin{split}
&
Q^{\ }_k (\theta,\phi)=
\left(
\begin{array}{ccc}
\mathbb{I}^{\ }_{2N^{\ }_{\mathrm{r}}-2} 
& 
0 
& 
0 
\\
0 
& 
q^{\ }_k (\theta,\phi) 
& 
0 
\\
0 
& 
0 
& 
-\mathbb{I}^{\ }_{2N^{\ }_{\mathrm{r}}-2}  
\\
\end{array}
\right),
\\
&
q^{\ }_k(\theta,\phi)=
\left(
\begin{array}{cc}
\cos\theta \mathbb{I}^{\ }_2 
& 
\sin\theta R^{\ }_{k}(\phi) 
\\
\sin\theta R^T_{k}(\phi) 
& 
-\cos\theta \mathbb{I}^{\ }_2 
\end{array}
\right),
\\
&
R^{\ }_k(\phi)
=
\left(
\begin{array}{cc}
\cos k\phi  
& 
\sin k\phi  
\\
-\sin k\phi 
& 
\cos k\phi   
\end{array}
\right),
\end{split}
\label{eq: Qk}
\end{equation}
labeled by the index $k\in\mathbb{Z}$.
One verifies that the Chern integer (Pruisken term)
$
\mathrm{Ch}[Q^{\ }_{k}]:=
\left(16\pi{i}
\right)^{-1}
\int^{\ }_{S^{2}} d^{2}r\,
\epsilon^{\ }_{\mu\nu}
\mathrm{tr}\,
\left[
Q^{\ }_{k}
\partial^{\ }_{\mu}
Q^{\ }_{k}
\partial^{\ }_{\nu}
Q^{\ }_{k}
\right]
$
vanishes for any $k\in\mathbb{Z}$.
This is expected as TRS holds.
On the other hand, we are going to argue that
\begin{equation}
\mathrm{sgn}\,
\mathrm{Pf}
\left(\widetilde{D}[Q^{\ }_{k  }]\right)=-
\mathrm{sgn}\,
\mathrm{Pf}
\left(\widetilde{D}[Q^{\ }_{k+1}]\right),
\quad
k\in\mathbb{Z}.
\label{eq: Pf for Qt}
\end{equation}
Any element
$Q\in G/H$
is also an element of the larger coset space
$\mathrm{U}(4N^{\ }_{\mathrm{r}})/
\mathrm{U}(2N^{\ }_{\mathrm{r}})\times
\mathrm{U}(2N^{\ }_{\mathrm{r}})$.
From this point of view,
the Chern integer 
is the signature of the second homotopy group
$
\pi^{\ }_{2}
\left(
\mathrm{U}(4N^{\ }_{\mathrm{r}})/
\mathrm{U}(2N^{\ }_{\mathrm{r}})\times
\mathrm{U}(2N^{\ }_{\mathrm{r}})
\right)
=\mathbb{Z}
$.
As the Chern integer is also the phase of the fermion determinant
of the Dirac kernel $D'[U]$ (through the chiral anomaly)
in the background of 
$
U \in 
\mathrm{U}(4N^{\ }_{\mathrm{r}})/
\mathrm{U}(2N^{\ }_{\mathrm{r}})\times
\mathrm{U}(2N^{\ }_{\mathrm{r}})$,
the fact that the Chern integer 
vanishes is a consequence of the positivity of
the determinant~(\ref{eq: Det D'}). 
While the Chern integer 
is blind to the second homotopy group%
~(\ref{eq: second homotopy group}),
the Pfaffian is sensitive to it.

\begin{figure}
  \begin{center}
  \includegraphics[width=8.2cm,clip]{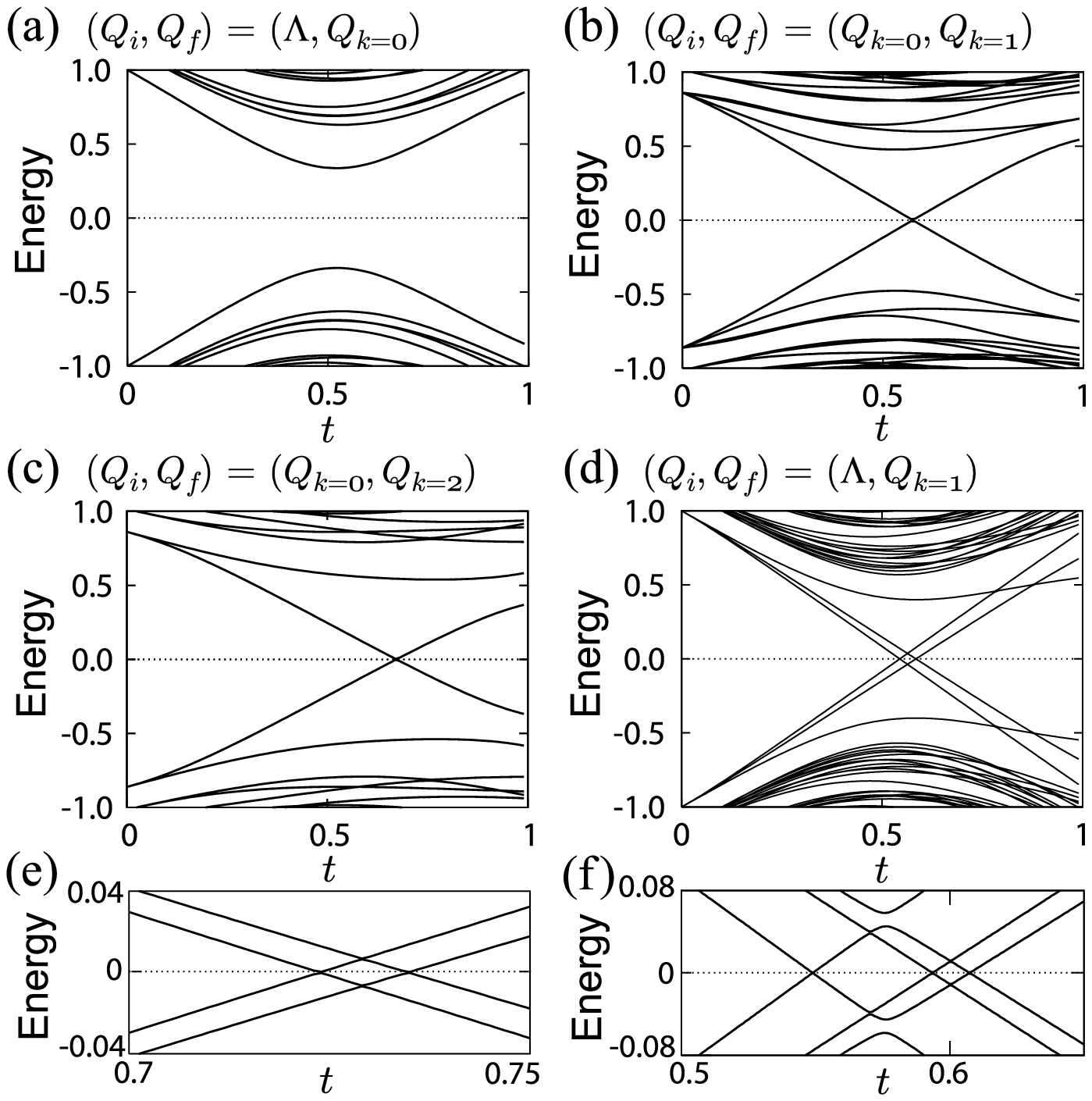} 
\caption{
\label{fig: spec_flow}
The energy eigenvalue spectrum 
in the vicinity of the band center 
for the Dirac kernel $D'[Q^{\ }_{t}]$,
Eq.~(\ref{eq: Dirac Kernel D'}),
is computed numerically as a function of the 
parameter $0\leq t\leq 1$
for
$(a^{\ }_{0} \Delta)^{-1}=1$ and $N=11$.
The field $Q^{\ }_{t}$
interpolates between 
$Q^{\ }_{i}$ when $t=0$
and
$Q^{\ }_{f}$ when $t=1$.
(e) and (f): 
Same as in (c) and (d), respectively,
except for the addition of a small perturbation
so as to lift any accidental or quasi degeneracies.
         }
\end{center}
\end{figure}

We give a ``numerical proof'' of Eq.~(\ref{eq: Pf for Qt}) 
in Fig.\ \ref{fig: spec_flow} by showing
4 evolutions of eigenvalues of the Dirac kernel%
~(\ref{eq: Dirac Kernel D'})
evaluated at
$
Q^{\ }_t:= 
(1-t)Q^{\ }_i 
+ 
t    Q^{\ }_f
$
as a function of $0\leq t\leq 1$.
Here, the initial, $Q^{\ }_{i}$, and final, $Q^{\ }_{f}$, 
field configurations belong to $G/H$~\cite{Witten82}.
Although 
$Q^{\ }_t$ 
is not a member of $G/H$ for $0<t<1$, 
it remains real-valued, symmetric, and traceless.
Consequently, the spectrum of $D'[Q^{\ }_t]$
is symmetric about the band center.
Configurations $Q^{\ }_i$ and $Q^{\ }_f$
have Pfaffians of opposite signs whenever an odd number
of level crossing occurs at the band center (``spectral flow'')
during the $t$-evolution 
of the Dirac kernel $D'[Q^{\ }_{t}]$. 
This is accompanied by
the closing of the spectral gap 
of $D'[Q^{\ }_{t}]$
by an odd number of pairs 
$(-\lambda'_{i},+\lambda'_{i})$
as $t$ interpolates between $0$ and $1$.
The spectral $t$-evolution
is obtained numerically using the regularization 
of the Dirac kernel $D'[Q^{\ }_{t}]$
in momentum space as described above 
Eq.~(\ref{eq: key identity between det})
\cite{note for numerics}.
We show with
Fig.\ \ref{fig: spec_flow}(a)
that $\Lambda$ (the uniform configuration) 
and $Q^{\ }_{k=0}$
belong to the same $\mathbb{Z}^{\ }_{2}$ 
topological sector as the spectral gap
never closes under the evolution of the spectrum:
$\mathrm{Pf}(\widetilde{D}[\Lambda])$
and 
$\mathrm{Pf}(\widetilde{D}[Q^{\ }_{k=0}])$
share the same sign.
We show with
Fig.\ \ref{fig: spec_flow}(b)
that $Q^{\ }_{k=0}$ and $Q^{\ }_{k=1}$
belong to different $\mathbb{Z}^{\ }_{2}$ 
topological sector as level crossing 
at the band center takes place for a single pair of levels:
$\mathrm{Pf}(\widetilde{D}[Q^{\ }_{k=0}])$
and 
$\mathrm{Pf}(\widetilde{D}[Q^{\ }_{k=1}])$
differ by their sign.
We show with
Figs.\ \ref{fig: spec_flow}(c) and (e)
that $Q^{\ }_{k=0}$ and $Q^{\ }_{k=2}$
belong to the same $\mathbb{Z}^{\ }_{2}$ 
topological sector as level crossing 
at the band center takes place for 2 pairs of levels
\cite{note: degeneracy}:
$\mathrm{Pf}(\widetilde{D}[Q^{\ }_{k=0}])$
and 
$\mathrm{Pf}(\widetilde{D}[Q^{\ }_{k=2}])$
share the same sign.
Finally, we show with
Fig.\ \ref{fig: spec_flow}(d) and (f)
that $\Lambda$ and $Q^{\ }_{k=1}$
belong to different topological sectors
as level crossing 
at the band center takes place for 3 pairs of levels
\cite{note: degeneracy}:
$\mathrm{Pf}(\widetilde{D}[\Lambda])$
and 
$\mathrm{Pf}(\widetilde{D}[Q^{\ }_{k=1}])$
have opposite signs.

From now on, we shall assign the topological quantum number 0 (1)
to all configurations belonging to the same homotopy class
as $Q^{\ }_{k}$ with $k$ even (odd).
If so, the effective action~(\ref{eq: def eff action})
can be approximated by~\cite{note: replica limit} 
\begin{equation}
Z^{\mathrm{topolo}}_{\mathrm{NL}\sigma\mathrm{M}}=
\int \mathcal{D}[Q]\,
(-1)^{n[Q]}\,
e^{-S[Q]}
\label{eq: NLS action with Z2 topology}
\end{equation}
where $S[Q]$ is the usual local action
for the NL$\sigma$M on 
$G/H$
and 
$n[Q]=0,1$ is the $\mathbb{Z}^{\ }_{2}$ 
topological quantum number of $Q$.
The topological term has its origin in the Pfaffian arising from
Majorana fermions, i.e., the global anomaly.
When the bare NL$\sigma$M coupling constant is small 
(when the bare conductivity is large), 
the effect of $n[Q]=0,1$ on 
the renormalization group flow is small.
Weak antilocalization, 
as it occurs in the absence of the topological term, 
is then expected.
The effect of $n[Q]=0,1$ is more important
in the strong coupling regime (when the bare conductivity is small)
\cite{Ando98}. 
Numerical simulations~\cite{Bardarson07,Nomura07} 
of the random Dirac Hamiltonian~(\ref{eq: single dirac hamiltonian})
suggest that the topological term makes
the two-dimensional symplectic stable metallic fixed point 
the only fixed point.

As is well known from lattice gauge theory \cite{Nielsen81},
the fermion doubling problem forbids the emergence of an odd
number of massless
spinors from two-dimensional tight-binding models with 
conserved electric charge, TRS, 
locality, and the translation symmetry of a regular lattice.
That is, the NL$\sigma$M~(\ref{eq: NLS action with Z2 topology})
cannot emerge from microscopic electronic models on 
the square or honeycomb lattices \cite{Obuse07}.
However, as pointed out by Ando and Suzuura \cite{Ando02},
graphene approximately realizes the symplectic symmetry 
if the potential range of static point 
impurities is much larger than the lattice spacing.

One way out of the fermion doubling problem is to consider
$d$-dimensional space as the boundary of $(d+1)$-dimensional space
\cite{Callan85,Frakin86}
and to study the physics of localization at this boundary.
In this context, the effect of disorder
on the two-dimensional boundary states
of three-dimensional $\mathbb{Z}^{\ }_2$ topological insulators
\cite{Moore06, Fu06, Roy06}
could yield microscopic realizations of the two-dimensional
symplectic class with a $\mathbb{Z}^{\ }_{2}$ topological term.
The situation is here similar to that in a quasi-one-dimensional
tight-binding model belonging to the symplectic symmetry class.
When the number of Kramers doublets propagating in the wire is even,
all states are exponentially localized \cite{Brouwer96}.
When it is odd, one Kramers doublet remains extended 
\cite{Takane04,Zirnbauer92}. 
The former case is always realized in a quasi-one-dimensional
tight-binding model \cite{Brouwer96},
whereas the latter case can be realized at 
the edges of  two-dimensional $\mathbb{Z}^{\ }_2$ topological insulators
\cite{Kane05}. 

C. M.,  H. O., and A. F. acknowledge hospitality of the Kavli Institute for 
Theoretical Physics at Santa Barbara, where this work was initiated.
This work was supported by Grant-in-Aid
for Scientific Research (Grant No.~16GS0219) from MEXT of Japan
and by the National Science Foundation
under Grant No.\ PHY99-07949.


\begin{thebibliography}{10}
\bibitem{note for numerics}
Without loss of generality,
we choose $N^{\ }_{\mathrm{r}}=1$ in which case
$Q^{\ }_{k}$ in Eq.~(\ref{eq: Qk}) reduces to the
$4\times4$ matrix $q^{\ }_{k}$.
We choose $N$ sufficiently large to avoid accidental 
level crossing.
Although, $a^{\ }_{0} \Delta$ 
is to be determined self-consistently,
we treat $\Delta$ as a free parameter.
The topologically non-trivial $Q$-field configurations
on the torus can be constructed in the same way as on the sphere.
We still call these configurations $Q^{\ }_k$ with $k\in\mathbb{Z}$.
One should also bare in mind that numerics cannot 
rigorously distinguish level crossing from avoided crossing.
\bibitem{note: degeneracy}
These levels can be accidentally or quasi degenerate for 
the choice~(\ref{eq: Qk})
[for example, $Q^{\ }_{k=0,2}$ have an accidental symmetry,
$Q^{\ }_{k=0,2}(\boldsymbol{r})=Q^{\ }_{k=0,2}(-\boldsymbol{r})$
on the torus].
Adding a small perturbation to $Q^{\ }_{i,f}$, as is done in 
Figs.\ \ref{fig: spec_flow}(e) and \ref{fig: spec_flow}(f),
lifts this (quasi) degeneracy.
\bibitem{note: replica limit} 
Observe first that the $4\times4$ block-matrix $q^{\ }_{k}$
in Eq.\ (\ref{eq: Qk})
is independent of $N^{\ }_{\mathrm{r}}$.
The global anomaly is thus expected to survive 
the replica limit $N^{\ }_{\mathrm{r}}\to0$.
Second, the derivation of the NL$\sigma$M that
requires a diffusive regime can be
achieved with the help of $N^{\ }_{f}$ flavors for the 
spinors. If so, the global anomaly is only possible 
when $N^{\ }_f$ is odd.
Third, the Dirac kinetic energy in 
Eq.~(\ref{eq: single dirac hamiltonian})
yields nothing but the $\widehat{\mathrm{so}}(4N^{\ }_{\mathrm{r}})^{\ }_{1}$ 
Wess-Zumino-Witten (WZW) critical point.
After disorder averaging, the disorder strength $g$ 
drives the system to the $G/H$ NL$\sigma$M.
\end{thebibliography}
\end{document}